\documentclass[aps,twocolumn,secnumarabic,nobalancelastpage,nofootinbibt]{revtex4}

\usepackage{graphics}      
\usepackage{graphicx}      
\usepackage{longtable}     
\usepackage{url}           
\usepackage{bm,color}            
\usepackage{amssymb, amsmath, enumerate, theorem, epsfig,subfigure,setspace}

\begin{document}

\title{Simulating the spectral gap with polariton graphs}

\author{Kirill Kalinin$^1$, Pavlos G. Lagoudakis$^{1,2}$,  and  Natalia G. Berloff$^{1,3}$ }
\email[correspondence address: ]{N.G.Berloff@damtp.cam.ac.uk}
\affiliation{$^1$Skolkovo Institute of Science and Technology Novaya St., 100, Skolkovo 143025, Russian Federation}
\affiliation{$^2$Department of Physics and  Astronomy, University of Southampton, Southampton, SO17 1BJ, United Kingdom}
\affiliation{$^3$Department of Applied Mathematics and Theoretical Physics, University of Cambridge, Cambridge CB3 0WA, United Kingdom }

\date{\today}

\begin{abstract}{We recently proposed polariton graphs as a novel platform for solving hard optimization problems that can be mapped into the $XY$ model. Here, we elucidate a relationship between the energy spectrum of the $XY$ Hamiltonian and the total number of condensed polariton particles. Using as a test-bed the hexagonal unit lattice we show that the lower energy states of the $XY$ Hamiltonian are faithfully reproduced by mean-field numerical simulations utilising the Ginzburg--Landau equation coupled to an exciton reservoir. Our study paves the way to simulating the spectral gap of the  XY model using polariton graphs.}
\end{abstract}

\maketitle
It is hard to identify a physical concept as important in condensed matter  physics   as the notion of the spectral gap. Phase transitions in the quantum many-body systems occur when the spectral gap vanishes, therefore phase diagrams crucially depend on its properties. Whereas, critical behaviour is associated with gapless systems, wherein long-range correlations are supported by low-energy excitations that behave as massless particles, non-critical behaviour is associated with gapped systems, wherein long-range correlations are prevented through massive low-energy excitations \cite{koma06}. In adiabatic quantum computation that can be as powerful as the usual circuit model for quantum computation \cite{aha07}, the spectral gap is a crucial quantity that defines the efficiency  of a quantum algorithm. Such algorithm is efficient  only if there exists a Hamiltonian path for which the minimal spectral gap is lower-bounded by an inverse-polynomial in the system size \cite{farhi00}. Finding a method that allows one to determine  whether a  quantum many-body Hamiltonian is gapped or not or even calculate the size of the gap is one of the fundamental questions  for condensed matter systems. Even  for some simple spin models on  1D and 2D lattices, there are famous outstanding problems on the existence of the spectral gap  \cite{gapexistance,gapexistance2}. Recently, it was rigorously proven that the spectral gap problem is undecidable; it is algorithmically impossible to say whether or not a general Hamiltonian is gapped or gapless \cite{cubitt1,cubitt2}. The implication is that one cannot study big enough, but still computable systems, detect the pattern and then extrapolate the results to a larger system that is not computable.

During the past decade one of the most promising applications of quantum information technology was in engineering a physical system that reproduces a many-body   Hamiltonian of interest; an analogue Hamiltonian simulator \cite{georgescu14}. Various physical systems have been proposed and realised to a various degree of scalability and efficiency. Ultracold atoms in optical lattices \cite{reviewUltracold,saffman,simon11,fermionic}, trapped ions \cite{kim10,lanyon}, photons \cite{northup}, superconducting q-bits \cite{corcoles}, network of optical parametric oscillators (OPOs) \cite{yamamoto11, yamamoto14}, and coupled lasers \cite{coupledlaser} are among the most promising systems proposed to overcome the limitations of the classical computation. The guidance on which spin Hamiltonians to emulate is given by the rigorously established  result on the  existence of  universal spin Hamiltonians.  All other classical n-vector  models with any range of interactions can be reproduced within such a model, and certain simple Hamiltonians such as the next-neighbour 2D Ising model on a square lattice with transverse fields are  universal \cite{cubittScience16}.

Recently we proposed and realised  an analogue Hamiltonian simulator on a polariton graph \cite{natmat17}. Polaritons are the composed light-matter bosonic quasi-particles  formed in the strong exciton-photon coupling regime in semiconductor microcavities \cite{weisbuch}. Due to bosonic stimulation  polaritons condense in the same quantum mechanical state \cite{Kasprzak,revKeelingBerloff, revCarusotto}.  Using spatial modulation  and   non-resonant optical excitation, polaritons can be made to condense at any location of a planar microcavity forming a two-dimensional graph of condensates \cite{keelingBerloffLattice}. When the coherence lifetime of polaritons exceeds the time of flight  between neighbouring sites (graph vertices), polaritons interactions lead to the development of phase relationships across the vertices \cite{tosi12, tosi13,ohadi16}.  As polaritons condense to the same quantum mechanical state, the phases of polaritons at the pumping sites become locked with  particular phase differences that  can be mapped into the spins of the $XY$ model.  Since the minimization of the $XY$ Hamiltonian is analogous to the maximization of the number particles in the condensate \cite{ohadi16}, at threshold density, polariton graphs condense with the phase/spin configurations that correspond to the ground state of the $XY$ Hamiltonian \cite{natmat17}.    The process of identifying the ground state of the $XY$ Hamiltonian through bosonic stimulation is very similar to that of coupled lasers \cite{coupledlaser, bosonicsti}.

In this letter, we establish that polariton graphs are not only capable of accurately finding the ground state of the $XY$ model, but also the low energy spectrum of the excited states, and therefore, can become an efficient tool for retrieving the spectral gap of $XY$ Hamiltonian. The energy landscape of the $XY$ Hamiltonian is set by the interaction strengths, $J_{ij}$, that depend on the pumping intensity and the graph geometry. Whereas at threshold density the system condenses at  the ground state of the corresponding $XY$ Hamiltonian \cite{natmat17}, we reason that for pump intensities {\it above threshold} the condensate occupies all stable energy states of the $XY$ Hamiltonian below the corresponding energy level of a given pump. We show that higher energy levels have progressively lower occupancy; we shall refer to the state with the second largest particle number as   ``the first excited state", where the difference in the number of particles between ground and first excited state represents the spectral gap of the $XY$ model.

The $XY$ model on different types of lattices, such as triangular \cite{Hauke_NewJPhys2010}, square \cite{Figueirido_PRB1990,Darradi_PRB2008, Read_PRL1991}, honeycomb \cite{Varney_PRL2011,Bishop_CondMat2012,Zhu_PRL2013}, is usually considered in terms of the frustration parameter $J_2/J_1$ representing the ratio of the strength of the next neighbour interactions, $J_2$, to the nearest neighbour interaction, $J_1$. A  system may exhibit different phase configurations depending on this  value: collinear ordering (i.e., antiferromagnetic ordering, N\'eel.I, classical order), the state of a quantum spin liquid (i.e., Bose metal), collinear ordering when two of the three nearest neighboring spins are antiparallel, and the other are parallel (i.e., N\'eel.II state, anti-N\'eel, collinear spin wave). 
The $XY$ model on a honeycomb lattice has attracted much attention of experimental and theoretical physicists, since a small number of neighbour interactions enhances quantum fluctuations, and therefore, it seems to be a promising system for obtaining spin liquid states. It was initially believed \cite{Varney_PRL2011} that for a simple $XY$ spin model, a specific spin-liquid ground state, a Bose liquid, appears for a particular range of the frustration parameter, while a surprising anti-ferromagnetic Ising phase was detected \cite{Zhu_PRL2013} for the same range by examining much larger lattices without finding any spin-liquid ground state. By considering models with second neighbour $J_1$-$J_2$ or even third neigbour $J_1$-$J_2$-$J_3$ interactions, possible symmetry breaking ground states were shown on a honeycomb lattice \cite{Mosadeq_CondMat2011,Bishop_CondMat2012}.


The total number of condensed polaritons in the system with $l$ equally pumped spots can be expressed as \cite{natmat17}:
\begin{equation}
N = \int |\psi(\textbf{r},t)|^2 d \textbf{r} \approx l N_0+ \sum_{i<j} J_{ij} (k_c,d_{ij}) \cos \theta_{ij},
\label{N_polaritons}
\end{equation}
where $\psi(\textbf{r},t)$ is  the condensate wavefunction, $N_0$ represents the number of polaritons of one isolated pumping spot. $J_{ij}$ stands for the interaction strength between polariton spots at positions ${\bf r}={\bf r}_i$ and ${\bf r}={\bf r}_j$, separated by the distance $d_{ij}=|{\bf r}_i-{\bf r}_j|$ with  outflow velocities $k_c$. Here $\theta_{ij} = \theta_i - \theta_j$ is the relative phase difference between the polaritons at  ${\bf r}_i$ and ${\bf r}_j$. From Eq. (\ref{N_polaritons}) we can define  the {\it particle mass residue} ${\cal M}=l N_0- N$ that represents the change in the number of particles in the system due to the interaction of the condensates among different pumping spots.   The expression for ${\cal M}$ from Eq. (\ref{N_polaritons}) approximates the definition of the $XY$ Hamiltonian, $H_{XY}=-\sum_{i<j} J_{ij} \cos \theta_{ij}$. The particle mass residue of above threshold polariton states corresponds to the energy spectra of the $XY$ model; the particle mass residue difference between ground and first excited state approximates the spectral gap of the $XY$ Hamiltonian.

\begin{figure}[h!]
	\centering
    \includegraphics[width=8.6cm]{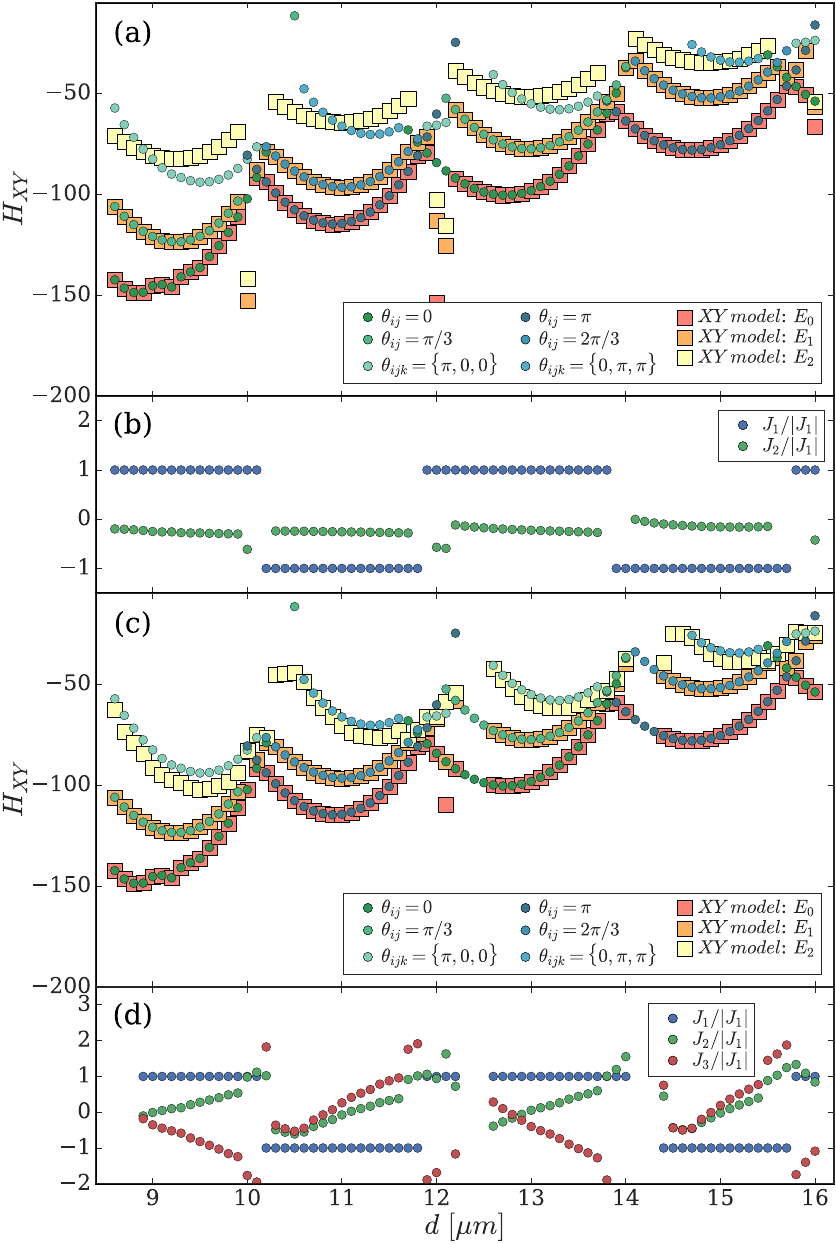}
	\caption{ (a,c) The lowest three energy levels of the XY model (squares) and the particle mass residues of polariton condensates (circles) as functions of the hexagon side, $d$. The particle mass residues are calculated by numerical integration of  Eqs. (\ref{GLE}-\ref{RateEq}) as discribed in the main text. The colour of the circles represents different phase configurations with the description given in the legend. The first three energy levels for (a) $J_1$-$J_2$ model (c) $J_1$-$J_2$-$J_3$ model, are found by the direct minimization of the XY Hamiltonian using the L-BFGS-B optimization and shown with red, orange, and yellow squares. Their phase configurations are similar to the phases shown with circles over which the squares are plotted. The coupling ratios with respect to the hexagon side $d$ found from Eqs. (\ref{GLE}-\ref{RateEq}), as described in the text, are plotted for (b) $J_1$-$J_2$ model and  (d) $J_1$-$J_2$-$J_3$ model (d).}
    \label{Fig1}
\end{figure}

We elucidate this argument by considering a hexagon unit cell with size $d=|{\bf r}_i - {\bf r}_{i+1}|$ and the pumping profile  $P = \sum_{i=1}^6 P_0 \exp(-\alpha |\textbf{r} - \textbf{r}_{i}|^2)$,  where $\alpha$ is the inverse width of the Gaussian. The particle mass residue becomes ${\cal M}=6 N_0 - N_{hex}$, where $N_{hex}$ is the number of particles in the hexagon of polariton condensates and the  XY Hamiltonian becomes
\begin{equation}
H_{XY}=- J_1\sum_{i=1}^6  \cos \theta_{ii+1} - J_2 \sum_{i=1}^6  \cos \theta_{ii+2}
 - J_3 \sum_{i=1}^3  \cos \theta_{ii+3},
 \label{N_hex}
\end{equation}
where the summation is cyclic in $i$ (e.g. $i+1$ is set to $1$ for $i=6$) and where we included all pairwise interactions between vertices. Experimentally,  the number of particles in the system and, therefore, the particle mass residues of the ground and then the lower excited states  are determined as the pumping intensity, $P_0$, approaches the condensation threshold from below and then exceeds it. This constitutes the speed up in comparison with the classical computer minimisation that requires an extensive search of the minima of an energy configuration of a high dimensionality fixed by the lattice size. For only six pumping spots we can compute the particle mass residues for the lower energy states numerically from the mean field equations based on the complex  Ginzburg-Landau equation (GLE)  written for the condensate wavefunction $\psi$ \cite{Wouters, Berloff}.
In Ref. \cite{natmat17} we established the set of parameters of the mean-field model of the polariton condensate that reproduces the experimental data across the full range of distances. In what follows, we use the same dimensionless model
\begin{eqnarray}	
	i   \frac{\partial \psi}{\partial t} &=& -  \left(1 - i \eta {\cal R} \right) \nabla^2\psi +  |\psi|^2 \psi+
 g {\cal R} \psi  \nonumber \\
	  &+&i\biggl( {\cal R} - \gamma \biggr) \psi, \label{GLE}\\
	  \frac{\partial \cal R}{\partial t} &=&  - \left( 1 + b |\psi|^2 \right) {\cal R} + P({\bf r}) ,
	\label{RateEq}
\end{eqnarray}
and the same set of the parameters and non-dimensionalization as in Ref. \cite{natmat17}. Here ${\cal R}$ is the dimensionless density profile of the exciton reservoir,  $g$  corresponds to the blue-shift due to interactions with non-condensed particles,  $\gamma$  represents the decay rates of condensed polaritons, $b$ is proportional to ratio of the rate at which the exciton reservoir feeds the condensate and the strength of effective polariton-polariton interaction, and $\eta$ is the energy relaxation coefficient specifying the rate at which gain decreases with increasing energy. The non-dimensionalization is chosen so that the unit length is 1$\mu$m. For a hexagon side $d$  between $8 \mu m $ and  $16 \mu m$, we find the stationary states by numerically  integrating Eqs. (\ref{GLE}-\ref{RateEq}) starting from hundred randomly distributed fields $\psi({\bf r},t=0)=\sum a_{\bf k} \exp(i {\bf k}\cdot{\bf r})$, where the phases of the complex amplitudes $a_{\bf k}$ are distributed uniformly on $[0,2\pi]$ \cite{berloffSvistunov}. The corresponding particle mass residues  are
shown  in Fig. \ref{Fig1}(a,c) with filled circles, where the different  colours correspond to various  phase differences between the hexagon vertices. For the parameters and distances considered, the polariton ground state has always $0$ (ferromagnetic (F)) or $\pi$ (antiferromagnetic  (AF)) phase differences. For F ground state the first and the second excited states  are always a single vortex with $\theta_{ij} = \pi/3$ and a spin wave with $\theta_{ijk} = \{\pi,0,0\}$, respectively, where $j$ and $k$ stand for short notation of adjacent condensates $i+1$ and $i+2$, respectively. For AF ground state, these are a double vortex with $ \theta_{ij} = 2\pi/3$ and a spin wave with $\theta_{ijk} = \{0,\pi,\pi\}$, respectively.

%
\begin{figure}[t!]
	\centering
    \includegraphics[width=8.6cm]{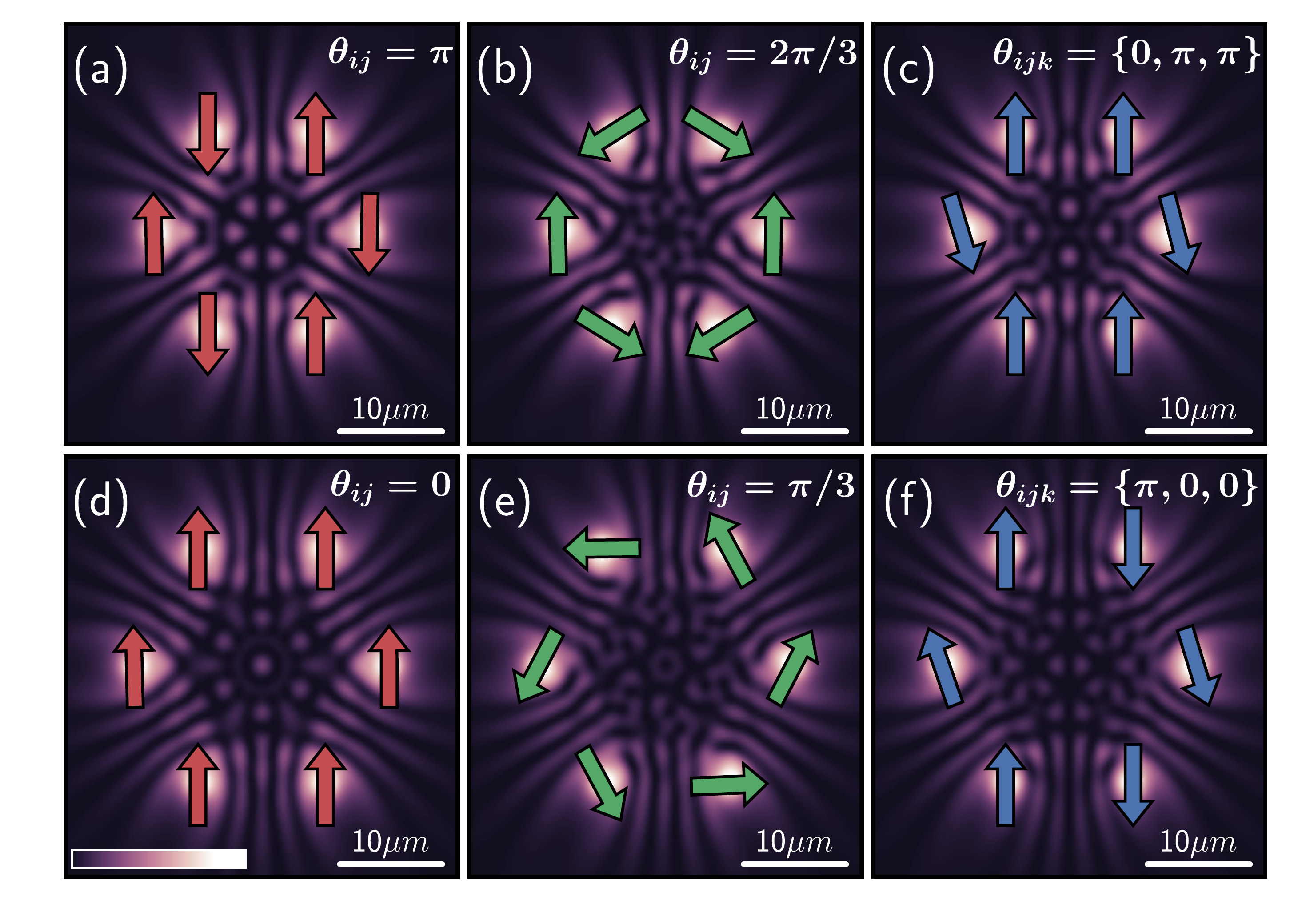}
	\caption{The polariton densites for the hexagons with the sides $d = 11 \mu m$ (a-c) and $d = 13 \mu m$ (d-f)  found by numerical integration of  Eqs. (\ref{GLE}-\ref{RateEq}). The first row shows (a) an AF ordering for the ground state (N\'eel's ordering), (b) a double vortex,  and  (c) a spin wave for the two lowest excited states (N\'eel.II state). The second row depicts (d) a F ordering, (e) a single vortex state, and (f) a different spin wave state (N\'eel.III state). The arrows correspond to the phases of the condensates. The indices $j$ and $k$ stand here for short notation of $i+1$ and $i+2$ neighbour condensates, respectively.}
    \label{Fig2}
\end{figure}

We can accurately estimate the coupling strengths for each hexagon side $d$ by solving the matrix equation ${\bf M}={\cal B} {\bf J}$, where
${\bf M}=[{\cal M}_{0},{\cal M}_{1}, {\cal M}_{2}]^T$, ${\bf J}=[J_1,J_2,J_3]^T$, and the matrix  ${\cal B}$ has elements $b_{mj}=q_j\sum_{i=1}^6 \cos\theta_{ii+j}^m$, $q_1=q_2=1$, $q_3=1/2$. Here, the elements of ${\bf M}$ are the particle number residues for the ground, the first and the second excited states of the polariton graph, respectively, and $m$ indexes the phases of the corresponding states.
First, we neglect $J_3$ interactions ($J_1$-$J_2$ model) and calculate the ratios of $J_1/|J_1|$ and $J_2/|J_1|$, that are shown in Figure \ref{Fig1}(b) with blue and green circles, respectively.
We use the obtained $J_1$ and $J_2$ for each $d$ to minimize the $XY$ Hamiltonian  by using the approximated Broyden-Fletcher-Goldfarb-Shanno algorithm (L-BFGS-B) \cite{L_BFGS_article1,L_BFGS_article2} starting from 1000 random initial conditions.
The resulting energies of the ground state and the two lowest excited states are denoted by filled squares in Fig. \ref{Fig1}(a) and show a good correspondence  between the GLE and the $XY$ model for the ground and the first excited states in terms of both the observed  phase configurations and the energy values. The phase configurations for the second excited states are generally predicted correctly and the energies are in a fair agreement. Figure \ref{Fig1}(c,d) shows the results for the solution of the full matrix equation  ( $J_1$-$J_2$-$J_3$ model), where a good agreement between all three states is illustrated.
The six distinct phase configurations that were observed for different hexagon sides $d$ in Fig. \ref{Fig1}(a,c) are shown in Figure \ref{Fig2} superimposed on the polariton densities. 

\begin{figure}[h]
	\centering
    \includegraphics[width=8.6cm]{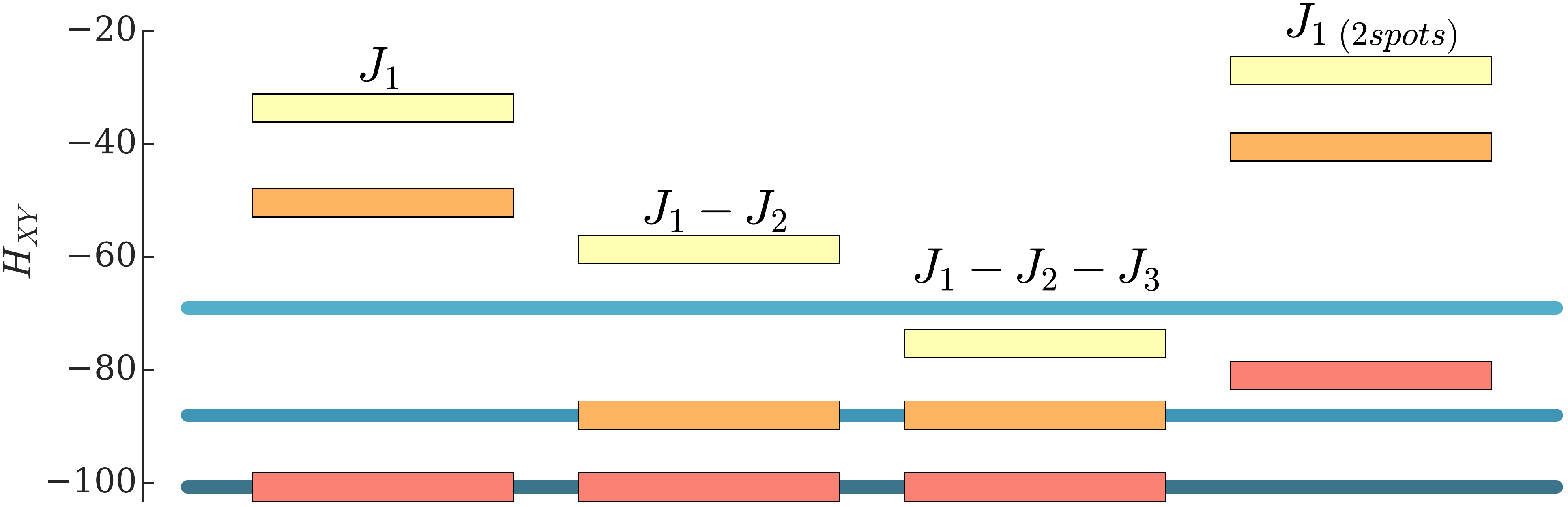}
	\caption{The comparison of the lowest energy levels of the four different $XY$ models, depicted with rectangles, with the polariton particle mass residues found from the GLE, depicted with blue solid lines, for the hexagon with the lattice constant $11.5 \mu m$. First column shows energy levels in case of $XY$ model including nearest neighbour interactions $J_1$, the second and the third columns include second neighbour interactions $J_1$-$J_2$ and the second and the third neighbour interactions $J_1$-$J_2$-$J_3$, respectively. These three models are based on the coupling strengths that are found through the analysis of the hexagon of polariton condensates. The last column shows the energy states of the $XY$ model with nearest neighbour interactions $J_{1 \ (2spots)}$ based on the couplings obtained from the analysis of the two isolated polariton condensates.}
    \label{Fig3}
\end{figure}
We summarize the differences between the energies and phase configurations of states found by a polariton graph and those predicted by the direct minimization of the $XY$ Hamiltonian in Fig. \ref{Fig3} for a particular hexagon side $d$. On this figure the polariton particle mass residues (blue lines) are compared  with the energy levels of the $XY$ model (squares) taking into account various coupling strengths: only $J_1$, $J_1$-$J_2$, $J_1$-$J_2$-$J_3$ as well as with $J_1$ coupling strengths obtained from the GLE model for two pumping spots only. The phase configurations (shown by various colours) coincide in all cases. The agreement between excited states becomes better when the further couplings $J_2$ and $J_3$ beyond the nearest neighbours are introduced.  The discrepancy between the energies of the ground states of the polariton particle mass residues and the $XY$ model, based on the coupling strenghs $J_1$ calculated for the two pumping spots, is contributed to the density enhancement from the remaining spots that change the outflow velocity $k_c$ and, therefore, the coupling strength. This implies that in order to use the coupling strengths found from pairwise interactions to construct the polariton graph one needs to find a way to compensate the density enhancements. We discuss the ways to achieve this elsewhere \cite{kalinin17density}.

In conclusion, we argue that  in polariton graphs the ``particle mass residues" of successive polariton states that occur with increasing excitation density above condensation threshold are a fair approximation of the $XY$ Hamiltonian's energy spectrum. We test our assumption in an hexagonal lattice unit; we calculate the phase configurations and spectrum of polariton condensates for a range of hexagonal lattice sizes using mean-field theory (GLE), and observe good agreement with the energy spectrum derived from the $XY$ model. Our study suggests that polariton graphs can be used as an efficient simulator for finding the spectral gap of the $XY$ spin model.




\end{document}